\documentclass[8pt]{llncs}
\usepackage{caption}
\usepackage{subcaption}
\usepackage{graphicx}
\usepackage{epstopdf}
\usepackage{pdfpages}
\usepackage{comment}
\usepackage{xspace}
\usepackage{color}
\usepackage{multirow}
\usepackage{bibnames}

    \usepackage{algpseudocode}
    \usepackage[frozencache=true,cachedir=.]{minted}
    \usepackage{algorithm}
    \usepackage{amsmath}
    \usepackage[capitalize]{cleveref}
    \usepackage{amsfonts}
    \usepackage{enumitem}

    \algrenewcommand\alglinenumber[1]{\tiny #1:}

    \newminted{c}{frame=single,fontsize=\tiny,linenos}
    \algnewcommand\algorithmicswitch{\textbf{switch}}
    \algnewcommand\algorithmiccase{\textbf{case}}
    \algnewcommand\algorithmicassume{\textbf{assume}}
    \algnewcommand\Assume[1]{\State \algorithmicassume(#1)}%

    \algdef{SE}[SWITCH]{Switch}{EndSwitch}[1]{\algorithmicswitch\ #1\ \algorithmicdo}{\algorithmicend\ \algorithmicswitch}%
    \algdef{SE}[CASE]{Case}{EndCase}[1]{\algorithmiccase\ #1}{\algorithmicend\ \algorithmiccase}%
    \algtext*{EndSwitch}%
    \algtext*{EndCase}%

\title{Input Validation with Symbolic Execution}
\author{Anay Mehrotra\and
Ayush Bansal\and
Awanish Pandey\and
Subhajit Roy}
\institute{Indian Institute of Technology Kanpur, India\\
\email{\{anay,ayushb,awpandey,subhajit\}@iitk.ac.in}} 
\newcommand{\toolname}{\textsc{InVaSion}\xspace}
\begin{document}
\maketitle
\begin{abstract}
				Symbolic execution has always been plagued by the inability to handle programs that require highly structured inputs. Most often, the symbolic execution engine gets overwhelmed by the sheer number of infeasible paths and fails to explore enough feasible paths to gain any respectable coverage. In this paper, we propose a system, \toolname, that attempts to solve this problem for \textit{forking-based} symbolic execution engines. We propose an input specification language (ISL) that is based on a finite-state automaton but includes guarded transitions, a set of registers and a set of commands to update the register states. We demonstrate that our language is expressive enough to handle complex input specifications, like the Tiff image format, while not requiring substantial human effort; even the Tiff image specification could be specified in our language with an automaton of about 35 states. \toolname translates the given program and the input specification into a non-deterministic program and uses symbolic execution to instantiate the non-determinism. This allows our tool to work with any \textit{forking-based} symbolic execution engine and with no requirement of any special theory solver. Over our set of benchmarks, on an average, \toolname was able to increase branch coverage from 24.97\% to 67.84\% over baseline KLEE.
\end{abstract}

\section{Introduction}\label{intro}
Symbolic Execution~\cite{king:1976} has emerged as a popular technique with increasing applications in program testing, program repair etc. Symbolic engines require the creation of \textit{symbolic variables} --- variables that are allowed to accept any value --- that can stand for the program inputs; the symbolic exploration of the program establishes a formal relation between the symbolic variables and the program outputs, which is exploited for applications like automated test generation and program repair. However, in many programs, not every possible stream of characters constitutes a valid input. In such cases, \textit{input modelling} i.e. modelling the precondition of programs that accept such structured inputs, has been identified as one of the biggest hurdles in deploying symbolic execution in the field.

\begin{figure}[h]
    \centering
    \begin{subfigure}{.55\textwidth}
        \centering
        \input{Code/totinfo.c.tex}
        \label{code:totinfo}
    \end{subfigure}%
    \begin{subfigure}{.44\textwidth}
        \centering
        \includegraphics[scale=0.6]{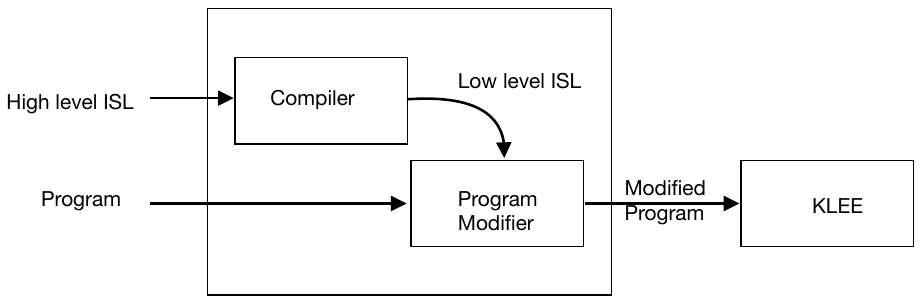}
        \label{layout}
    \end{subfigure}
    \caption{(a) Tot\_info code showing EXIT\_FAILURE on invalid inputs; (b) Basic Layout of \toolname}
\end{figure}
Let us illustrate the problem of input modelling with an example: Fig. 1a shows a snippet of code from the \textit{Tot\_info} program, a popular benchmark from the \textit{Siemens} suite~\cite{sir}. The program accepts a sequence of lines from the input file such that:
\begin{itemize}
	\item Each line is a space separated string of digits (which will be split in line 6 and 16);
	\item The integers corresponding to the first two strings correspond to the number of rows and columns (line 6);
	\item The product of these numbers correspond to the size of a subsequent matrix to be populated; the size of this matrix must be below MAXTBL (line 10);
    \item On subsequent lines, it should be followed by \textit{exactly} one (or more) space separated string of numbers (to be converted into long integers), and the number of these strings must be equal to the matrix size.
\end{itemize}

This is a fairly complex precondition! Failure to meet any of these constraints terminates the program with an error code EXIT\_FAILURE (at lines 8, 12 and 18). Running the program with KLEE~\cite{klee:2008} results in no successful path. Overall, even within a time budget of 1800s, KLEE is only able to reach a line coverage of 29\% (which is mostly code for format checking).

To circumvent this problem, the developer is required to insert a sufficient set of \textit{assume} statements to embed the input model. This requires a good understanding of the code (not just the input format) so that these assume statements can be inserted at the relevant places.

In this work, we design a domain-specific language, Input Specification Language (ISL), suited for such input modelling. We also build a tool, \toolname, that when fed with an input specification (in our DSL) and an input program, invokes a symbolic execution engine in a manner such that it \textit{only} explores program paths corresponding to those in the input specification. Our DSL is structured as a finite-state machine --- with a guarded transition function, finite number of registers to record additional state information and ability to execute commands that manipulate the state --- that only accepts valid inputs. Our experiments demonstrate that even complex input specifications can be modelled with simple finite-state machines: for example, \cref{hisl:totinfo} shows the input specification for \textit{Tot\_info} (we explain it in \cref{sec:overview}). Even for complex input formats like for programs that read Tiff images, an ample input specification of Tiff could be accommodated within a finite-state machine with only 34 to 38 states, while improving the coverage of these programs by more than 50\% on an average. Over a set of benchmark programs, spanning from \texttt{Qsort, FFT, Tot\_info to TiffSplit, Tiff2rgba}, we achieve more than 50\% increase in coverage with finite-state machines of 3 to 36 states. \\
\hspace*{1cm} This problem has been found to be important and has been studied previously~\cite{raghavan,loop,cab,Majumdar,god,lin,caba}. However, all these proposals have been in the context of \textit{instrumentation-based} symbolic execution engines~\cite{DART}. To the best of our knowledge, ours is the first work for \textit{forking-based} symbolic execution engine. The following are the differences and challenges addressed in our work: \\
\hspace*{0.5cm} Instrumentation based symbolic execution engines use a random execution (to begin with), perform a concrete execution on the engine, while constructing the path condition as the concrete execution progresses. Finally, at the end of a path, it constructs a new path by flipping one of the branch conditionals; the check for the input precondition happens at this point. For example, Godefried~\cite{god} uses a special solver that conjuncts the membership check in the input grammar with the path condition, to generate a new feasible path (to be explored). \\
\hspace*{0.5cm} On the other hand, for forking-based engines, the engine primarily does a symbolic interpretation, forking into new states at branches. If the above strategy is adopted, we would be required to invoke the expensive, special \textit{context-free constraint solver} at each branch, thereby slowing down execution. We solve this problem in the following way:
\begin{itemize}
				\item \textbf{specification translation} We provide an algorithm for translating the input specification into a fragment of first-order logic with linear inequalities and uninterpreted functions;
				\item \textbf{input slicing} Instead of handling the complete input specification along each symbolic path, we fragment the specification into a disjoint set of constraints. This allows for each symbolic path to work with ``lighter" preconditions, leading to faster solver times and an overall high coverage in a given time budget.
				\item \textbf{symbolic execution for non-determinism} We translate the input program to a non-deterministic program, where each non-deterministic run models a fragment of possible inputs. We exploit the symbolic execution for implementing this non-determinism. In fact, the program from \toolname can be run on \textbf{any, off-the-shelf, symbolic execution engine} and requires \textbf{no special theory-solving capability}.
\end{itemize}

Additionally, we propose a new input description language---essentially an enhanced automaton---that makes it easy to describe input patterns that require semantic interpretation of formatted strings (like ``...the input should be a list of length of \texttt{n} integers, where \texttt{n} is the first element in the list").

The following are our contributions:

\begin{itemize}
				\item We design a new domain-specific language that is suitable for the input modeling. Our DSL is structured as a finite-state machine with a guarded transition function, finite registers and the ability to execute commands that manipulate the state. Our experiments demonstrate that even complex input specifications can be captured in our language with a small number of states. In the future, we propose to provide \textit{input specification libraries} to abstract away common specifications such as mini-FSAs that can be invoked from the primary master FSA.

	\item We design an algorithm that allows a symbolic execution engine to traverse paths corresponding to those specified in this specification. We instantiate our ideas into a tool, \toolname.

        \item We evaluate \toolname on a set of programs from Siemens testsuite~\cite{sir} and MiBenchmarks~\cite{Mi}, achieving more than 50\% increase in code coverage with small FSAs spanning 3 to 37 states.
\end{itemize}

\section{Overview} \label{sec:overview}

Fig. 1b shows the basic layout of \toolname. Let us illustrate how our tool operates on the \texttt{Tot\_info} program. (\cref{code:totinfo})

\begin{figure}[h]
    \centering
    \begin{subfigure}{.49\textwidth}
        \centering
        \includegraphics[scale=0.65]{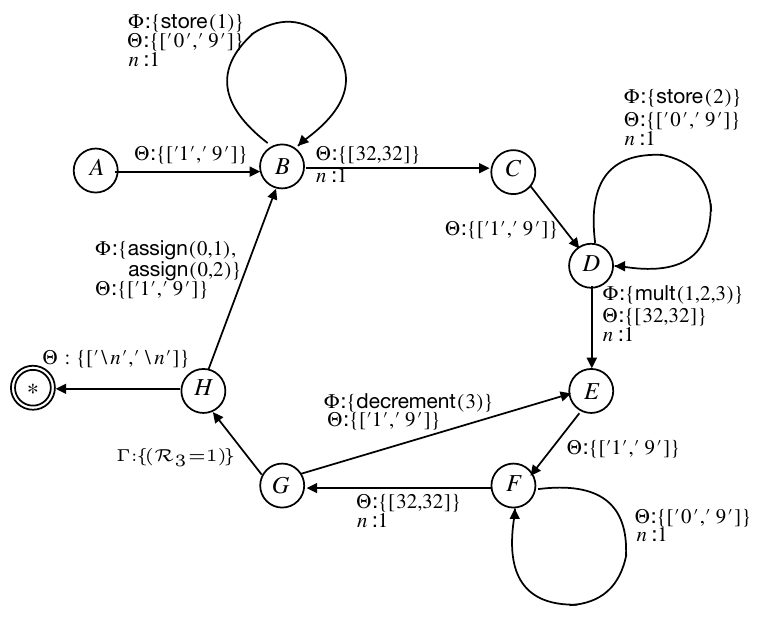}
        \caption{High level FSA}
        \label{hisl:totinfo}
    \end{subfigure}%
    \begin{subfigure}{.49\textwidth}
        \centering
        \includegraphics[scale=0.65]{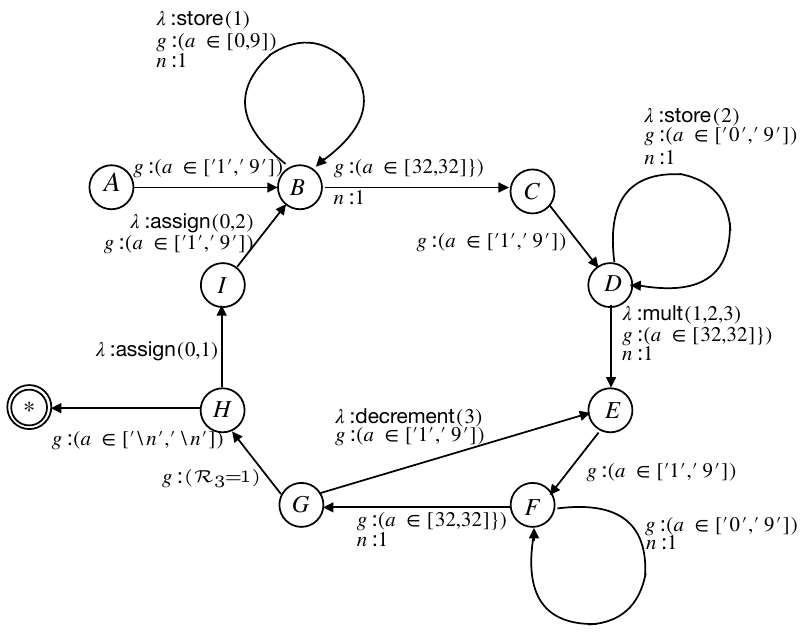}
        \caption{Low level FSA}
        \label{lisl:totinfo}
    \end{subfigure}
    \caption{FSA for Tot\_info}
    \label{isl:totinfo}
\end{figure}

    To begin with, the user provides the input specification as shown in \cref{hisl:totinfo}.
    The $8$ state finite-state automata (FSA), starts at state \texttt{A} and accepts the input at state \texttt{*}. Broadly, we can partition the automata into $4$ sections, states \texttt{A} and \texttt{B}, states \texttt{C} and \texttt{D}, loop of states  \texttt{E-F-G}, and the singleton state \texttt{H}. The first and the second section accept single integers representing the number of rows and columns respectively. $\mathcal{R}_1$ ($\mathcal{R}_1$ denotes the values stored in \texttt{Register[1]}) stores the number of rows and is updated at state \texttt{B}. Similarly, $\mathcal{R}_2$ corresponds to the number of columns and is updated at state \texttt{D}. Following this the transitions from state \texttt{D} to state \texttt{E} stores the number of entries in the matrix in $\mathcal{R}_3$.
    The loop of states \texttt{E-F-G} then accepts each entry of the matrix, decrementing the value in $\mathcal{R}_3$ after accepting each entry and exiting when the value of $\mathcal{R}_3$ is 1. Notice that on encountering a \texttt{null} (premature termination of input) or any invalid character the automaton rejects the input immediately. Once all entries have been accepted by the automaton, if the input is terminated by a newline, the automaton accepts it; otherwise, if there are further integers in the input, the automaton transitions to section 1. While transitioning to section 1, the automaton resets $\mathcal{R}_1$ and $\mathcal{R}_2$, initialising a new matrix. If the input has any invalid characters, the automaton rejects it. \\
    Next, the \textit{ISL compiler} in \toolname compiles this high-level ISL into a low-level ISL with fewer primitives as shown in \cref{lisl:totinfo}.
    The low-level ISL has a similar structure to the high level ISL. The difference being an additional state \texttt{I} in the low-level ISL. The low-level ISL allows at most 1 command per transition. This forces us to split the transition from \texttt{H} to \texttt{B}, and introduce an additional node.
Next, the \textit{input constraint generator} in \toolname non-deterministically selects valid (symbolic) inputs in accordance to the provided input specification and invokes the symbolic execution engine on it. Note that \toolname does not enforce a concrete input, but only a set of logical constraints that forces the input to remain valid. For example, it could force a string to belong to a string of digits corresponding to a ``small" integer (i.e., say $[1-9][0-9]^n$, where $n\leq 4$), but still maintains it as symbolic; the exploration through KLEE may concretize it, for example while generating tests.

    \begin{figure}[h]
        \includegraphics[scale=0.8]{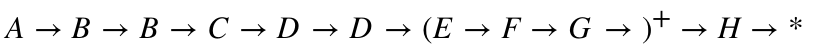}
        \centering
        \caption{A Path taken by symbolic execution engine}
        \label{path:totinfo}
        \includegraphics[scale=0.8]{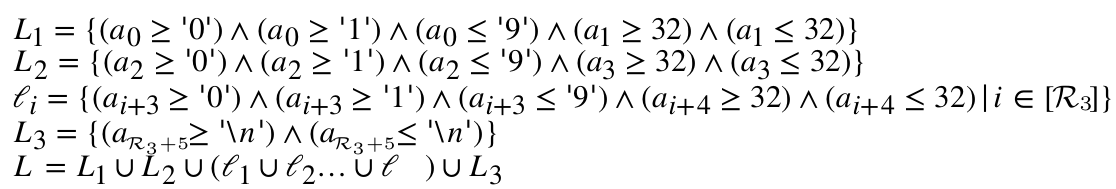}
        \centering
        \caption{Constraint Set generated for Tot\_info}
        \label{const:totinfo}
    \end{figure}

    Every path in the FSA corresponds to a class of inputs for which KLEE is invoked (in fact, our implementation is run inside KLEE to avoid the overhead of invoking KLEE multiple times): \cref{const:totinfo} shows the logical constraints generated for the input path shown in \cref{path:totinfo} in the FSA (\cref{lisl:totinfo}) and its subsequent symbolic execution tree in KLEE. At present, these constraint chains (FSA paths) are picked as per the KLEE path selection heuristics. In the future, we are looking to allow users to provide scheduling hints such that certain input classes are \textit{scheduled} ahead of others. This can be particularly useful if the developers suspect certain input classes to be more susceptible to errors (like larger strings are more likely to invoke buffer overflows than shorter strings).

\section{Input Specification Language (ISL)}
    \subsection{Low-level ISL} \label{ssec:lisl} \label{lisl}

        \toolname uses the ISL compiler to compile the high-level ISL (discussed in \cref{ssec:hisl}) to a low level model.
        The low-level ISL is essentially a non-deterministic finite-state machine that takes a string as input, and decides its validity based on its reachability to the final state.
	The input stream (i.e. character stream) can be seen as a \textit{one-way, read-only} input tape (i.e. the automaton can only read the input once moving left to right).
	Note that this specification requires semantic interpretation of parts of the input string (like the first two whitespaces separated string of digits); such specifications are difficult to specify using other forms like context-free grammars.

    \noindent
    Formally, our finite-state machine for the low-level ISL can be described as a 9-tuple:
	$$\mathcal{M} = (\mathcal{Q}, \Sigma, \mathcal{R}, \Lambda, \mathcal{G}, \Delta, s, f, r)$$
        where
        \begin{itemize}
            \item $\mathcal{Q}$ is a finite set of FSA states.
            \item $\Sigma$ is a finite set of input characters.
	    \item $\mathcal{R}$ is a finite set of registers; these registers provide additional (finite) memory for storing auxilliary state information while analyzing the input string.
            \item $\Lambda$ is a finite set of commands; these commands accept constants or registers as arguments to modify the state of the registers;
	    \item $\mathcal{G}$ is a boolean guard for transitions
	    \item $\Delta$ is a transition relation $\mathcal{Q} \times \mathcal{G} \times \Lambda \times \mathbb{Z} \rightarrow \mathcal{Q}$;
		    A transition $(q_s, g, \lambda, n, q_d) \in~\Delta$
		    \begin{itemize}
			     \item $q_s, q_d \in \mathcal{Q}$ are the source and destination states;
			    \item $\lambda \in \Lambda$ is a \textit{command} or register-update function that describes how the auxilliary state information in the registers need to be updated;
			    \item $g \in \mathcal{G}$ is a boolean \textit{guard} expression; the transition is enabled only if this guard is true (or equivalently, the transition immediately moves to the rejection state if the guard evaluates to false);
			    \item $n \in \mathbb{Z}, n\geq0$ denotes the number of characters on the tape to be consumed (or equivalently, increment to the input tape pointer).
      		    \end{itemize}
            \item $s \in \mathcal{Q}$ is the start state.
            \item $f \in \mathcal{Q}$ is the accept or final state.
	    \item $r \in \mathcal{Q}$ is the reject (trap) state.
        \end{itemize}

\noindent
	The language for the transition guards is a conjunction of a comparison on a register and a range-check for the input character, with an option of omitting any (or both) of the two checks:
$\mathcal{G} ::= \mathcal{S}_1 \land \mathcal{S}_2$ \\
$\mathcal{S}_1 ::= true \ |\  \mathcal{R}_i \oplus \mathbb{Z}$\\
$\mathcal{S}_2 ::= true \ |\ a \ominus [c_1, c_2]$\\
where
                        \begin{itemize}
			    \item $\oplus \in \{=,\neq,<,\leq,>,\geq\}$, and $\ominus \in \{\in, \notin\}$.
                            \item $\mathcal{R}_i$ is the value stored in $i_{th}$ register.
                            \item $\mathbb{Z}$ is an integer constant
                            \item $a$ is the character read and $c_1,c_2$ are characters constants, $a,c_1,c_2 \in \Sigma$.
                        \end{itemize}
        \noindent
	A transition relation $(q_s \in \mathcal{Q}, g \in \mathcal{G}, \lambda \in \Lambda, n\in \mathbb{Z}, q_d \in \mathcal{Q}) \in \Delta$ is handled as follows: given that the current state of the FSA is $(q_s, \rho_s, p_s)$, where $q_s$ is the state of the FSA, $\rho_s$ is the state of the registers and $p_s$ is the pointer to input character, the FSA transits to a new state $(q_t, \rho_t, p_t)$, where

    \begin{equation}
    (q_t, \rho_t, p_t) =
        \begin{cases}
            (q_d, \lambda(\rho_s), p_s+n) & \quad \text{if }g(\rho_s, p_s) = true\\
            (r, \_, \_)             & \quad \text{if }g(\rho_s, p_s) = false
        \end{cases} \label{eq:cons}
    \end{equation}

	If the guard function $g$ evaluates to true, the FSA transits to the destination state $q_d$ with a new state of the registers computed by applying the commands $\lambda$ on the current state of the registers, and incrementing the input tape pointer by $n$ positions; else the FSA transits to a REJECT (trap) state.
    Following are the examples of commands which can be used:
    \begin{itemize}
        \item \textit{add\_i} and \textit{mult\_i} take 3 arguments: source register, integer constant and destination register.
        \item \textit{add}, \textit{sub} and \textit{mult} take 3 arguments: 2 source registers, and a destination register
        \item \textit{assign} takes 2 arguments: source register and destination register
        \item \textit{increment} and \textit{decrement} take 1 argument: source register.
        \item \textit{store} takes 2 arguments along with $a$ (character which has been read): source register and destination register (adds ASCII($a$) to source register and stores in destination).
        \item Users can define their own C functions with \texttt{void} return type such that these functions take registers and integer constants as arguments.
    \end{itemize}

    Our input description language is powerful enough for the targeted benchmarks. Our language can be made expressive enough to handle context-free languages by adding a stack along with the set of registers. 

    \subsection{High-level Syntax} \label{ssec:hisl}

    The input specifications of the programs having formatted input can be very simple (like Qsort program) or highly complex (like Tiff programs). The complexity of input specifications makes it difficult for someone to precisely design the low level non-deterministic finite state machine (discussed in \cref{ssec:lisl}). \newline
    \toolname takes a high level ISL as input and compiles it to the low level ISL (discussed in \cref{ssec:compile}).
    The high level ISL is very similar to the low level ISL except in the definition of transitions. The transitions in this model can contain a huge amount of information compared to the previously discussed model and thus, shortens the representation of programs having highly complex input specifications. \newline
    Formally, the definition of the transition in the high level ISL will differ from low level ISL as per following:
    \begin{itemize}
        \item Instead of a single command $\lambda \in \Lambda$, a composition (sequence) of commands $\Phi = \{\lambda_1, \lambda_2, ...\}$ such that $\lambda_i \in \Lambda$ can be provided; $\Phi$ can also be empty (i.e. no command has to be executed).
        \item The boolean guard $g \in G$, is replaced by the following $2$ expressions:
            \begin{itemize}
                \item $\Gamma$ is a boolean guard defined on the registers such that it is a conjunction of terms which are comparisons on registers i.e. $((\mathcal{R}_i\oplus \mathbb{Z}_i) \land (\mathcal{R}_j \oplus \mathbb{Z}_j) \land ...)$; this conjunction will be used to construct $S_1$ for $g$. $\Gamma$ can also be $true$ (i.e. empty) implying no register comparisons.
                \item $\Theta$ is either a set of range of characters (can be inclusive or exclusive) or a set of sequence of characters (strings); this set will be used to construct $S_2$ for $g$.
            \end{itemize}
    \end{itemize}
\subsection{Compiling high-level ISL to low-level ISL} \label{ssec:compile}
\toolname takes the high-level ISL (discussed in \cref{ssec:hisl}) as input and compiles it to the low-level ISL (discussed in \cref{ssec:lisl}).
The transitions of high-level ISL are broken down and converted to transitions compatible with low-level ISL, in general increasing the number of states. \newline

    \begin{figure}[h]
        \centering
        \begin{subfigure}{.23\textwidth}
            \centering
            \includegraphics[scale=0.65]{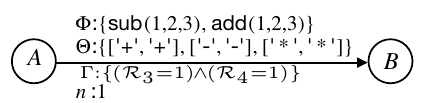}
            \caption{High level ISL}
            \label{hisl:example1}
        \end{subfigure}
        \begin{subfigure}{.42\textwidth}
            \centering
        	\includegraphics[scale=0.5]{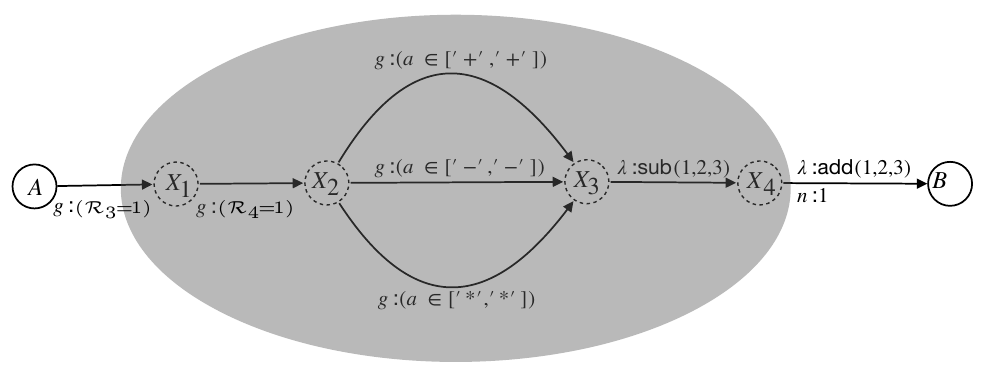}
        	\caption{Expanded low-level ISL}
        	\label{isl:inter}
        \end{subfigure}
        \begin{subfigure}{.32\textwidth}
            \centering
            \includegraphics[scale=0.5]{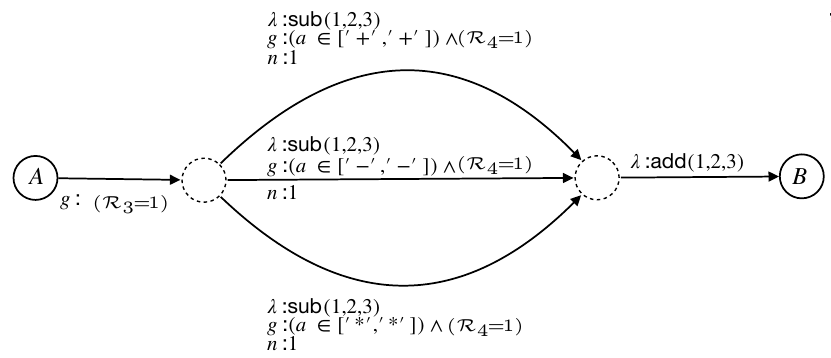}
            \caption{Low level ISL}
            \label{lisl:example1}
        \end{subfigure}%
        \caption{Example A of ISL (set of range of characters)}
    \end{figure}

    \begin{figure}[h]
        \centering
        \begin{subfigure}{.39\textwidth}
            \centering
            \includegraphics[scale=0.8]{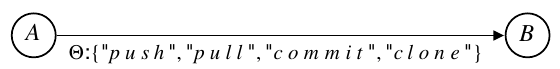}
            \caption{High level ISL}
            \label{hisl:example2}
        \end{subfigure}
        \begin{subfigure}{.59\textwidth}
            \includegraphics[scale=0.6]{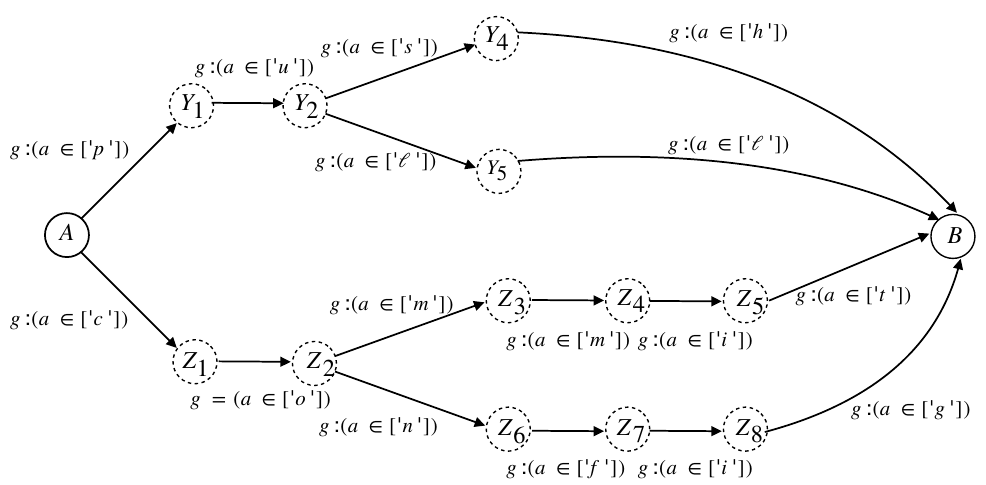}
            \centering
            \caption{Low level ISL}
            \label{lisl:example2}
        \end{subfigure}%
        \caption{Example B of ISL (set of strings)}
    \end{figure}

We illustrate this translation with an example for the high level transition from state A to state B in \cref{hisl:example1} and \cref{hisl:example2}. 
The transitions are~$(A, \Gamma=(R_3=1\land R_4=1),\Theta=(a\in[($`$+$'$,$`$+$'$),($`$-$'$,$`$-$'$),($`$*$'$,$`$*$'$)]),\Phi = (\text{sub},\text{add}), n, B)$ and $(A,\Theta=(``push",``pull"$$,$ $``commit"$$,$ $``config"), n, B)$. We unfold each of these complex transitions into a sequence of simpler transitions as supported by our low-level ISL. To achieve this we first unfold the guards, i.e., the register checks and character checks, followed by the composition of commands.
\begin{enumerate}
				\item Since, commands can alter the state of the registers,  the commands can commence only after the \textit{register checks} are completed. For example, the \textit{register check} $\mathcal{R}_3=1\land \mathcal{R}_4=1$ is expanded into the register checks $\mathcal{R}_3=1$ and $ \mathcal{R}_4=1$ from state  $A$ to $X_1$ and from $X_1$ to $X_2$ respectively.
				\item Following \textit{register checks}; \textit{character checks} are expanded. Depending on $\Theta$, a set strings or a set of character ranges are expanded. Let us first consider a set of strings.

                    This is done by combining all the strings which share a prefix and then accepting this prefix through common transitions, i.e., we build a \textit{trie} from the strings. This reduces the total number of states in the ISL. For example, in \cref{hisl:example2}, strings $``push"$ and $``pull"$ which share the prefix $``pu"$, also share path from state $A$ to $Y_2$ in the low-level ISL.

                    In the other case, a set of characters ranges is unfolded; it is done by unfolding the transition into multiple transitions between the same states, each accepting a single character from the set. For example, in \cref{isl:inter} the character set $\Theta=\{($`$+$'$,$`$+$'$),($`$-$'$,$`$-$'$),($`$*$'$,$`$*$'$)\}$ is expanded into 3 parallel transitions between states $X_2$ and $X_3$.
    \item Composition of commands is expanded next. For example, in \cref{isl:inter} the composition $\Phi = \text{sub}(1,2,3)\cdot\text{add}(1,2,3)$ is expanded into individual operations along transitions from $X_3$ to $X_4$ and from $X_4$ to $B$.
    \item Finally, the input pointer is incremented in accordance to the value $n$, of the transition. For example, in the \cref{isl:inter} the pointer is incremented on a transition between $X_4$ and $B$ by setting $n=1$.
\end{enumerate}

We attempt to overlap the above operations while staying within the syntax of low-level ISL to achieve smaller transition systems. For example, the shaded regions in the figure are merged into fewer transitions: transitions from $X_1$ to $X_2$ and $X_3$ to $X_4$ are merged into transitions from $X_2$ to $X_3$.

\section{Algorithm}
    \toolname which is based on KLEE using its forking mechanism at each branch for implementing the non-deterministic selection of transitions and the \textit{klee\_assume} statements for updating the constraint list. It accepts an input string only if \textit{any} of the possible transition sequences through the automaton (corresponding to the input ISL) reaches the \textit{accept} state.

    \begin{algorithm}[h]
      \caption{Automaton for the FSA}
      \scriptsize
      \begin{algorithmic}[1]
	      \Procedure{\toolname}{FSA $\mathcal{M}$, program $\mathcal{P}$}\Comment{Start state and pointer to start of string}
	    \State $\rho \gets \vec{0}$ \Comment{Initialize registers}
	    \State ${ip} \gets 0$ \Comment{Initialize input pointer}
	    \State $q_s \gets s$ \Comment{Initialize current state to initial state}
	    \State $\mathcal{L}\gets \{\}$ \Comment{Constrain set}
            \While{$q_s$ is not final state}
                \State Non-deterministically pick a transition $(q_s, g, \lambda, n, q_d) \in \Delta$, initiating from state $q_s$
                    \If{register check is enabled}
                    \Assume{$\mathcal{R}\oplus \mathbb{Z}$}
								    \State add constraint {$\mathcal{R}\oplus \mathbb{Z}$} to $\mathcal{L}$
                    \EndIf
                    \If{character check is enabled}
												 \Assume{$input[ip] \, \ominus \, [c_1, c_2]$}
                                 \If{$\ominus$ is $\in$}
                                    \State add constraint {$[(input[ip] \geq c_1) \land (input[ip] \leq c_2)]$} to $\mathcal{L}$
                                 \Else
                                    \State add constraint {$[(input[ip] < c_1) \lor (input[ip] > c_2)]$} to $\mathcal{L}$
                                 \EndIf
                    \EndIf
		    \State $\rho \gets \lambda{(\rho)}$ \Comment{execute the register-update function}
                \State $ip \gets ip  + n$\Comment{move the input pointer}
                \State $q_s\gets q_d$\Comment{move to the next state}
            \EndWhile
	    \State  Invoke symbolic execution on $\mathcal{P}$ with input tape reset and precondition $\mathcal{L}$
        \EndProcedure
      \end{algorithmic}
    \end{algorithm}

    After the selection of a transition if the guard is satisfied, the command is executed and transition system moves to the next state after incrementing input stream; else transition system moves to reject state (eq.~\eqref{eq:cons}~in~\cref{lisl}).
    
    Let us consider the \texttt{Tot\_info} program (Fig. 1a) and its FSA as shown in~\cref{lisl:totinfo}.
    The steps in the execution of the algorithm for the accepted path mentioned in \cref{path:totinfo} will be as follows:

    We denote $a_{ip}$ as the input character read at the current input position($ip$); if the command provided in a transition is $\emptyset$ then no command is executed and if the tape increment is $0$ then $ip$ is unchanged.

    \begin{enumerate}[label=\textbf{S.\arabic*}]
        \item Variables are initialized as mentioned in lines 2-5; $\rho$ (register values) are set to $0$; $ip$ (input pointer) is set to the start of the string i.e. $0$; $q_s$ (current state of automaton) is set to the state $A$ (start state of the FSA).
        \item \label{algo:1}At this point the available transition is $A\to B$. The algorithm chooses the transition $(A, a\in[$`$1$',`$9$'$], \emptyset, 0, B)$ (at line 7).

				\begin{itemize}
								\item For this transition, the guard is $(a_0\in[$`$1$',`$9$'$])$, and the constraint $[(a_0 \geq $`$1$'$) \land~(a_0 \leq $`$9$'$)]$ gets added to the constraint set $\mathcal{L}$ (at line 15);
								\item The transition system moves to state $B$, by appropriately updating $q_s$ (at line 22). 
				\end{itemize}
            \item \label{algo:2}At state $B$, from the available transitions $(B\to B)$ and $(B\to C)$; let us assume the algorithm (non-deterministically) chooses the transition $(B, a\in[$`$0$',`$9$'$], store(1), 1, B)$ (at line 7).
        \begin{itemize}
            \item For this transition, the guard is $(a_0\in[$`$0$',`$9$'$])$, and $[(a_0 \geq $`$0$'$) \land~(a_0 \leq $`$9$'$)]$ is added to $\mathcal{L}$ (at line 15).
            \item $store(1)$ command is executed (at line 20) changing the current state of registers by storing the input character's value in register $1$.
            \item $ip$ is incremented from $0$ by tape increment value which is $1$ (at line 21) and transition system moves to state $B$ (line 22).
        \end{itemize}
    \item \label{algo:3}Now, from the available transitions $(B\to B)$ and $(B\to C)$ let us assume the algorithm chooses transition ($B, a\in[32,32], \emptyset, 1, C$) (line 7).
        \begin{itemize}
            \item For current transition, the guard is $(a_1\in[32,32])$, and $[(a_1 \geq 32) \land (a_1 \leq 32)]$ is added to $\mathcal{L}$ (at line 15).
            \item The input pointer is incremented to $2$ (tape increment value is $1$).
            \item The transition system moves to state $C$, by appropriately updating $q_s$ (at line 22).
        \end{itemize}
    \item At this state, available transition is $(C\to D)$. The algorithm chooses the transition $(C, a\in[$`$1$',`$9$'$], \emptyset, 0, D)$ (at line 7); this step is similar to \ref{algo:1}, the constraint $[(a_2 \geq $`$1$'$) \land (a_2 \leq $`$9$'$)]$ gets added to set $\mathcal{L}$ and transition system moves to state $D$.
    \item At state $D$, from the available transitions $(D\to D)$ and $(D\to E)$; let us assume the algorithm chooses the transition $(D, a\in[$`$0$',`$9$'$], store(2), 1, D)$ (at line 7); similar to \ref{algo:2}, the constraint $[(a_2 \geq $`$0$'$) \land (a_2 \leq $`$9$'$)]$ is added to $\mathcal{L}$, $store(2)$ command is executed and $ip$ is incremented to $3$.
    \item Now, from the available transitions $(D\to D)$ and $(D\to E)$ let us assume the algorithm chooses the transition ($D, a\in[32,32], mult(1,2,3), 1, E$) (line 7).
        \begin{itemize}
            \item For current transition, the guard is $(a_3\in[32,32])$, and $[(a_3 \geq 32) \land (a_3 \leq 32)]$ is added to $\mathcal{L}$ (at line 15).
            \item $mult(1,2,3)$ command is executed (at line 20) setting $\mathcal{R}_3 = \mathcal{R}_1 \times \mathcal{R}_2$; $ip$, is incremented from $3$ to $4$ (tape increment is $1$).
            \item the transition system moves to state $E$, by appropriately updating $q_s$ (at line 22).
        \end{itemize}
    \item At this state, available transition is $(E\to F)$. The algorithm chooses the transition $(E, a\in[$`$1$',`$9$'$], \emptyset, 0, F)$ (at line 7); this step is similar to \ref{algo:1}, the constraint $[(a_4 \geq $`$1$'$) \land (a_4 \leq $`$9$'$)]$ gets added to set $\mathcal{L}$ and transition system moves to state $F$.
        \item At state $F$, from the available transitions $(F\to F)$ and $(F\to G)$; let us assume the algorithm chooses the transition $(F, a\in[$`$0$',`$9$'$], \emptyset, 1, F)$ (at~line~7).
        \begin{itemize}
            \item For this transition, the guard is $(a_4\in[$`$0$',`$9$'$])$, and $[(a_4 \geq $`$0$'$) \land (a_4 \leq $`$9$'$)]$ is added to $\mathcal{L}$ (at line 15).
            \item $ip$ is incremented from $4$ by tape increment value which $1$ (at line 21) and transition system moves to state $F$ (line 22).
        \end{itemize}
    \item Now, from the available transitions $(F\to F)$ and $(F\to G)$ let us assume the algorithm chooses the transition ($F, a\in[32,32], \emptyset, 1, G$) (line 7); similar to \ref{algo:3}, the constraint $[(a_5 \geq 32) \land (a_5 \leq 32)]$ is added to $\mathcal{L}$, the input pointer, $ip$, is incremented from $5$ to $6$ (tape increment value is $1$) and the transition system moves to state $E$.
    \item Now, the available transitions are $(G\to E)$ and $(G\to H)$; since the input values are symbolic, the value $\mathcal{R}_3$ is also symbolic; let us assume that algorithm picks transition ($G, a\in[$`$1$',`$9$'$], decrement(3), 1, E$) till $\mathcal{R}_3 = 1$.
	    \begin{itemize}
	    				\item For this transition, the guard is $(a_6\in[$`$1$',`$9$'$])$, and $[(a_6 \geq $`$1$'$) \land (a_6 \leq $`$9$'$)]$ gets added to the constraint set $\mathcal{L}$ (at line 15);
                        \item $decrement(3)$ command decrements the value of $\mathcal{R}_3$ (line 20).
	    				\item The transition system moves to state $E$, by appropriately updating $q_s$ (at line 22). 
	    \end{itemize}
    \item Now, at state $E$ let us assume the algorithm non-deterministically follows the same path again while $\mathcal{R}_3 \neq 1$ at state $G$.
    \item At state $G$, when $\mathcal{R}_3 = 1$, let us assume the algorithm picks the transition ($G, \mathcal{R}_3 = 1, \emptyset, 0, H$)
        \begin{itemize}
    	    			\item For this transition, the guard is $\mathcal{R}_3 = 1$, and $\mathcal{R}_3 = 1$ gets added to the constraint set $\mathcal{L}$ (at line 15);
	    				\item The transition system moves to state $H$, by appropriately updating $q_s$ (at line 22). 
        \end{itemize}
    \item Finally at state $H$, the available are transitions are $(H\to *)$ and $(H\to I)$; let us assume the algorithm picks the transition $(H, a\in[$`$\backslash n$'$], \emptyset, 0, *)$
	    \begin{itemize}
            \item For this transition, the guard is $(a_{i+6}\in[$`$\backslash n$'$])$ ($i$ is the value of~$\mathcal{R}_1\times~\mathcal{R}_2$), and $[(a_{i+6} \geq $`$\backslash n$'$) \land (a_{i+6} \leq $`$\backslash n$'$)]$ gets added to the constraint set $\mathcal{L}$ (at line 15);
	    				\item The transition system moves to state $F$, by appropriately updating $q_s$ (at line 22). 
	    \end{itemize}
    \item Now, the automaton has reached the final state $*$, it will exit the while loop (line 6) and invoke symbolic execution on program $\mathcal{P}$ with input tape reset back and preconditions of $\mathcal{L}$ imposed on input.

    \end{enumerate}

\section{Experiments}{
     Our experiments performed on 4 Core Machine with 6 GB RAM. We selected benchmarks from siemens~\cite{sir} and Mi Benchmarks~\cite{Mi} which require formatted input.

     Our experiments were designed to answer the following research questions:
     \begin{itemize}
	     \item[RQ1]{\bf Expressiveness} Does our DSL possess adequate expressive power to model a variety of input modeling problems?
	     \item[RQ2]{\bf Human Effort} What is the human effort required to construct the input models? Can complex input constraints be specified using small FSAs?
	     \item[RQ3]{\bf Total Coverage} Given good input specifications, is \toolname capable of significantly improving the total coverage?
	     \item[RQ4]{\bf Cumulative Coverage} Does the rate of coverage increase appreciably with the number of tests/time?
     \end{itemize}

     \subsection{RQ1: Expressiveness}
     	\toolname is indeed capable of describing a variety of input modeling scenarios. We attempted to model a set of 10 input modelling scenarios from popular benchmark suites.
     We discuss the input modeling challenges for each of our benchmarks below:

     \begin{itemize}
         \item \textbf{Tiff} All files are organized into $2$ \textit{compulsory} sections namely \textit{Image File Header} (IFH), and \textit{Image File Directory} (IFD) and an optional section namely the image bitmap data. A \texttt{Tiff} file can contain multiple images (no limit); $1$ IFD and $1$ bitmap per image. The only part of a Tiff file that has a fixed location is the IFH (first $8$ bytes of every Tiff file), all other data in Tiff file is represented by IFD, an IFD along with its bitmap forms a Tiff \textit{subfile}, the location of both IFD and bitmap can be variable.
            Each IFD contains one or more data structures called \textit{tag}. Each \textit{tag} is $12$ bytes that contain information about a specific property of the bitmapped data. They are found in contiguous groups and the identifiers of these tags (integers) occur in ascending order in the IFD. Each tag defined in Tiff has specific values for the $12$ bytes allocated to it and in the current version of Tiff there are about $44$ pre-defined tags out of which about $11$ commonly occur in Tiff files.
             Tiff has a reputation for being a complicated format due to variable locations and very specific organisation of tags.

    \item \textbf{Tot\_info}
    Already discussed in \cref{intro}.

    \item \textbf{Schedule}
        and \textbf{Schedule2} are online priority schedulers, meaning they accept a stream of input representing current tasks and updates the itinerary accordingly. Both programs accept the same input format. The programs expect $3$ integers as command line arguments. The three arguments represent the number of tasks with low, medium and high priority respectively.
    Following this, the programs accept multiple lines, where each line is a space separated string of integers and floats.
    The first integer, $i\in [7]$, in each line represents one of the 7 operations allowed by the programs. Depending on the operation this may be  followed by integers and/or floats, as specified in the following.
    \begin{itemize}
      \item \texttt{new\_job}, $i=1$: Followed by an integer $j\in [3]$, representing the priority of the new task. Here, $1,2,3$ correspond to low, medium and high priority respectively. 
      \item \texttt{upgrade\_prio}, $i=2$: Followed by an integer $j\in [2]$ and a float $r$. $j$ represents the priority of the task and $r$ is used to calculate the task number. Given the size of priority queue $j$, $\texttt{size}(j)$, the program updates the priority of the $k$-th task in the queue $j$ to $j+1$, where $k=\lfloor \texttt{r}\cdot \texttt{size}(j)\rfloor$.
      \item \texttt{block}, $i=3$: Not followed by an integer or float. It moves the current task to the blocked queue.
      \item \texttt{unblock}, $i=4$: Followed by a float, $r$, which is used to calculate a task number in the blocked queue. Given the size of the blocked queue, $\texttt{size}$, it removes the $k$-th task in the blocked queue, where $k=\lfloor r\cdot \texttt{size}\rfloor$.
      \item \texttt{quantum\_expire}, $i=5$: Not followed by an integer or float. It moves the current task to the end of its priority queue.
      \item \texttt{finish}, $i=6$: Not followed by an integer or float. It finishes the current task and removes it from the priority queue.
      \item \texttt{flush}, $i=7$: Not followed by an integer or float. It clears all priority queues.
    \end{itemize}
    \item \textbf{BasicMath}: is used to solve a single cubic equation. The  program accepts \textit{exactly} 4 space separated integers each of which represent the coefficients of the cubic equation.
    \item \textbf{FFT}: accepts a string of digits as the first argument; following this; a string of digits \textit{specifically} representing a power of $2$ as the second argument. 
    \item \textbf{Qsort}
    Accepts a single line of space separated string of integers.
    %
    \item \textbf{Binary Search}
    The program accepts a single line of space separated strings of digits, where each string represents an integer.
    The first integer, say $n$, corresponds to the size of the array. The next integer represents the key $k$, to be searched in the array. After this, the program accepts \textit{exactly} $n$ more integers, each representing a value $v$, in the array.
\end{itemize}
     \subsection{RQ2: Human Effort}
     Table~\ref{tab:nodes} shows the number of states in the FSA for each benchmark. In most cases, our High-level FSA has about 10 states even for complex specifications like Schedule and Tot\_info. Even the specification of the Tiff image format could be done in about 35 states.
		 The table also shows the blowup on using the compilation to the low-level ISL, thereby saving significant human effort.

    \begin{minipage}{\textwidth}
      \centering
      \begin{minipage}[b]{0.69\textwidth}
        \centering
        \includegraphics[width=\linewidth]{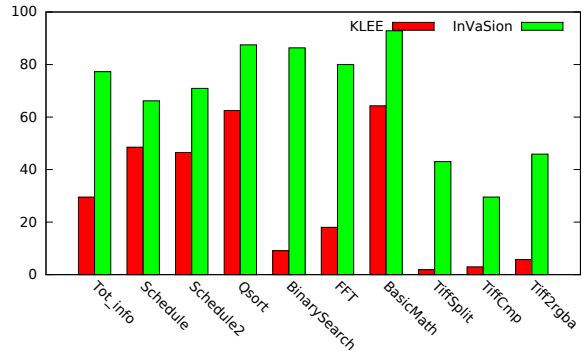}
          \captionof{figure}{Branch Coverage achieved by the test-cases generated by KLEE and \toolname} \label{fig:cov}
      \end{minipage}
      \hfill
      \begin{minipage}[b]{0.29\textwidth}
        \centering
        \scalebox{0.95}{\begin{tabular}{|l|l|l|}
            \hline
              Benchmark & HL & LL \\ \hline
              Tot\_info & 8 & 9\\ \hline
              FFT & 4 & 42 \\ \hline
              Qsort & 3 & 3 \\ \hline
              BasicMath & 5 & 6 \\ \hline
              TiffSplit & 35 & 122 \\ \hline
              Tiff2rgba & 36 & 125 \\ \hline
              TiffCmp & 36 & 125 \\ \hline
              Binary Search & 13 & 15 \\ \hline
              Schedule & 11 & 18 \\ \hline
              Schedule2 & 11 & 18 \\ \hline
        \end{tabular}}
          \captionof{table}{Number of nodes in High-level and Low-level ISL} \label{tab:nodes}
      \end{minipage}
    \end{minipage}
     \subsection{RQ3: Total Coverage}
		 Fig~\ref{fig:cov} shows the total coverage. In many benchmarks, we obtain significant improvement in coverage; for example, in \texttt{Tot\_info}, our coverage improves from 29.25\% to 77.27\% (with a small FSA of 8 nodes).
	On an average, our benchmarks gain an increase in branch coverage of 48\%  across all the benchmarks (from 24.97\% to 67.84\% on an average, i.e. an improvement of 171.6\%).

     \subsection{RQ4: Cumulative Coverage}
     Fig~\ref{fig:timeVsCov} shows how \toolname enables a faster increase in coverage of relevant paths with respect to KLEE. Across the benchmarks, by the time we attain 25\% coverage, KLEE attains 21.25\%, when we attain 50\% coverage.

    \begin{figure}
            \begin{subfigure}[b]{.24\textwidth}
                    \centering
                    \includegraphics[width=\linewidth]{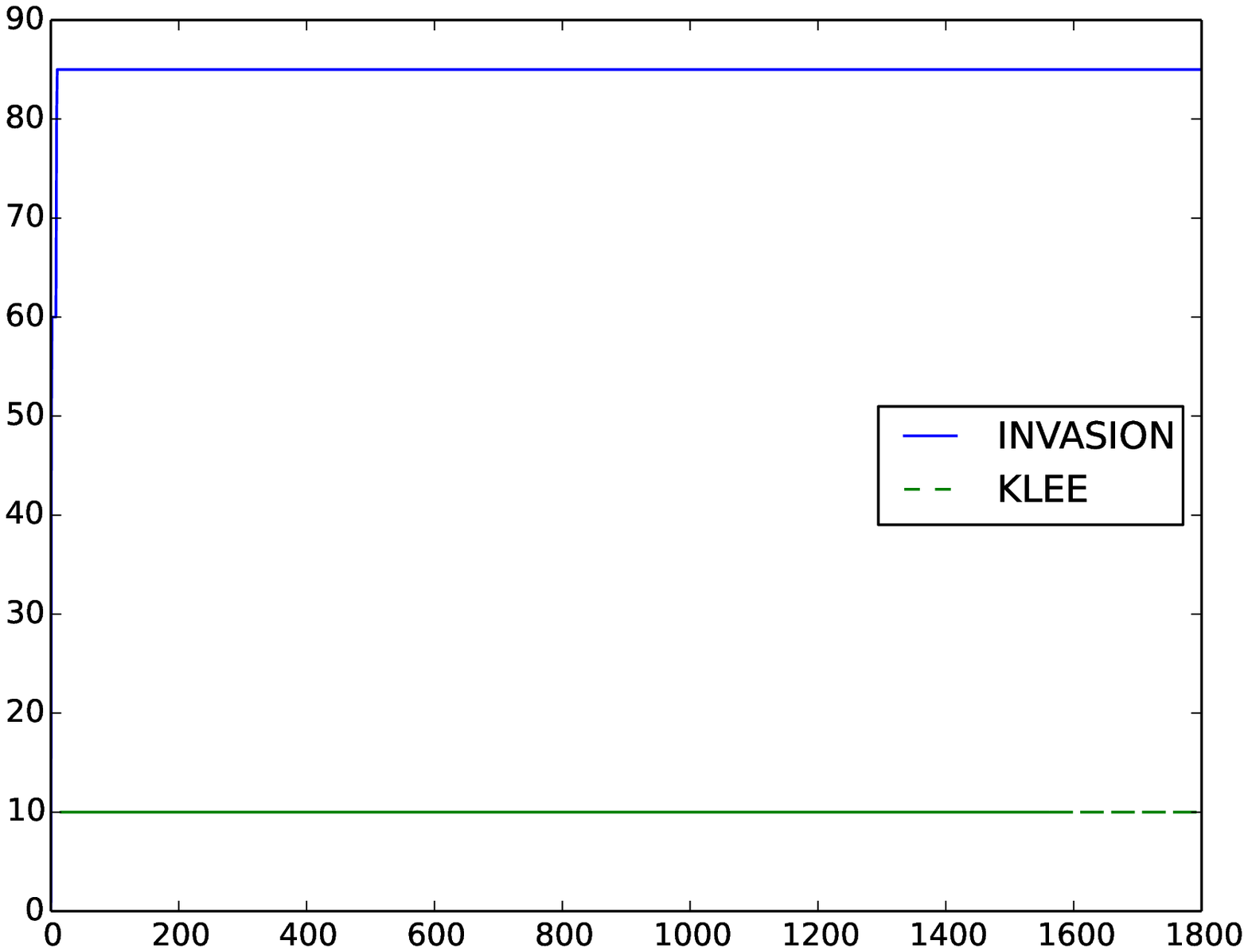}
                    \caption{BinarySearch}
                    \label{fig:gull}
            \end{subfigure}
            \begin{subfigure}[b]{0.24\textwidth}
                    \centering
                    \includegraphics[width=\linewidth]{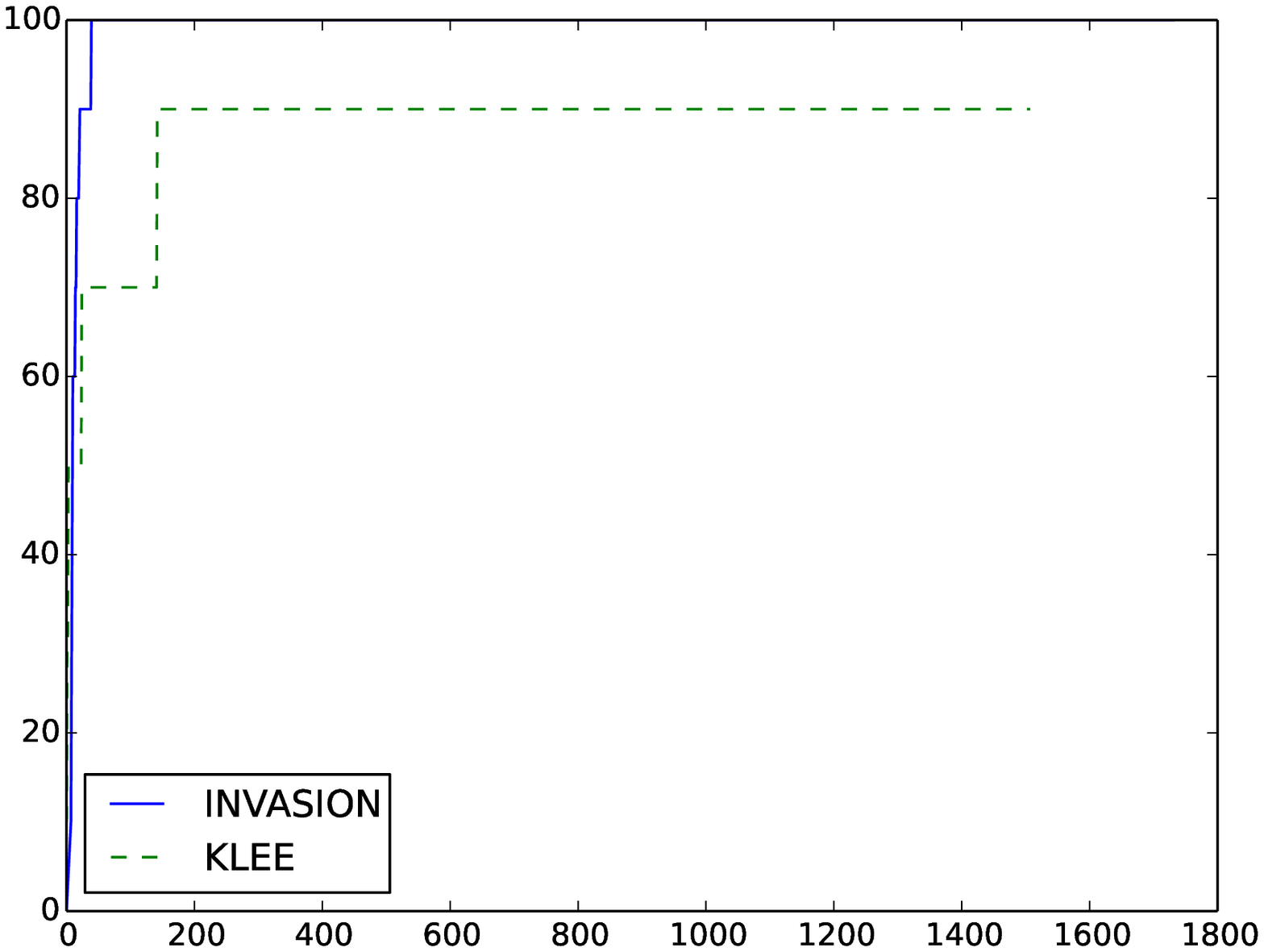}
                    \caption{BasicMath}
                    \label{fig:gull2}
            \end{subfigure}
            \begin{subfigure}[b]{0.24\textwidth}
                    \centering
                    \includegraphics[width=\linewidth]{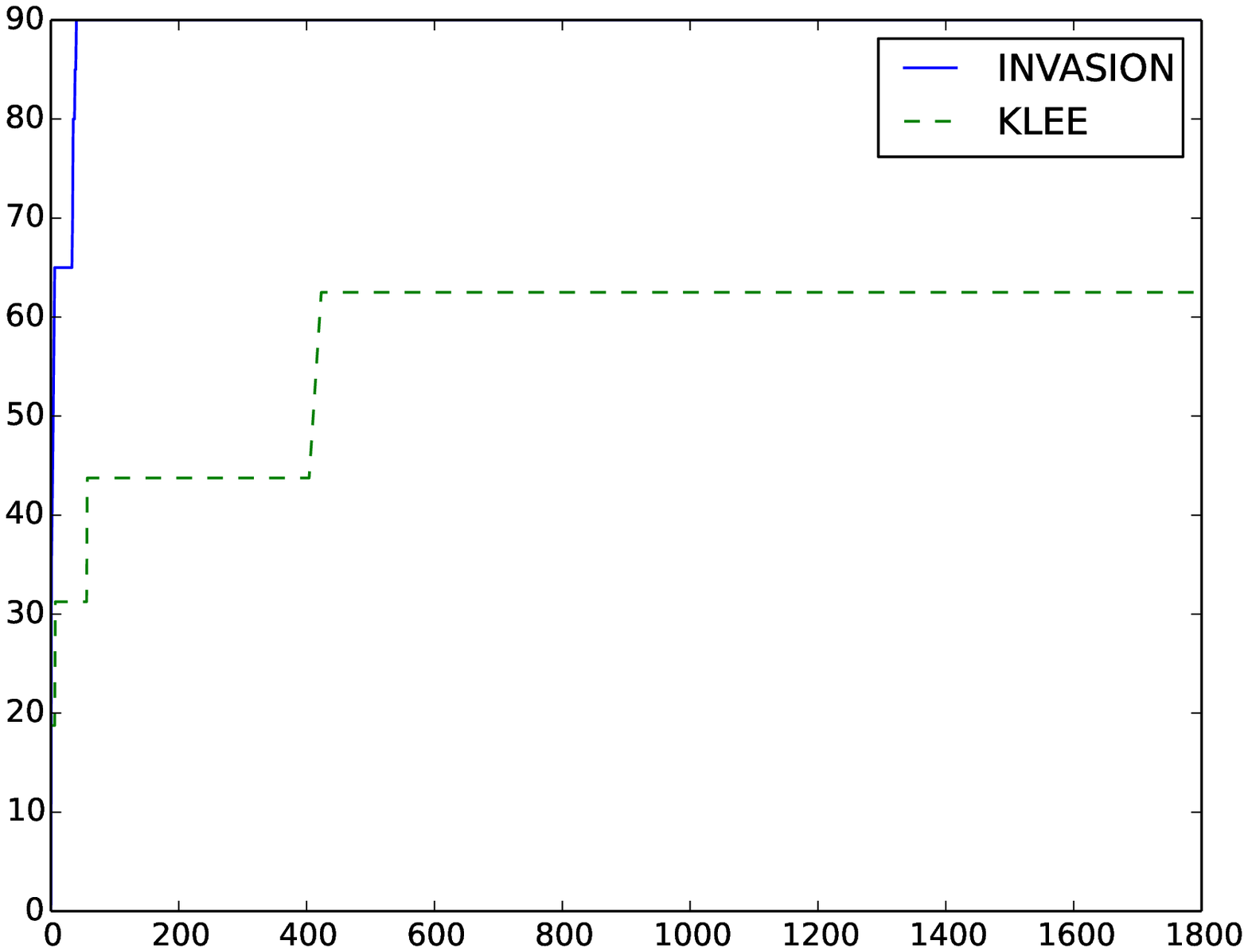}
                    \caption{QSort}
                    \label{fig:tiger}
            \end{subfigure}
            \begin{subfigure}[b]{0.24\textwidth}
                    \centering
                    \includegraphics[width=\linewidth]{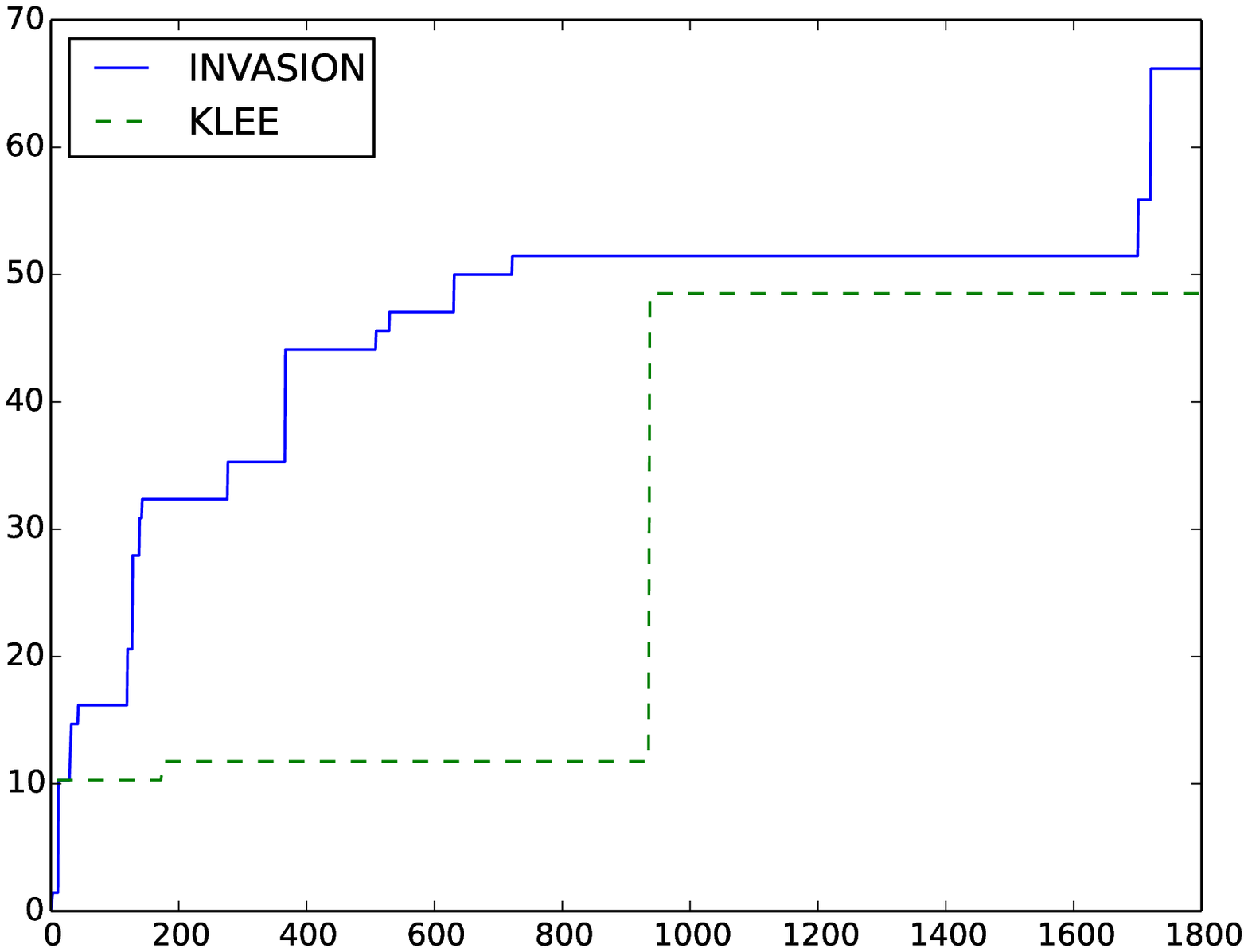}
                    \caption{Schedule   }
                    \label{fig:mouse}
            \end{subfigure}
            \begin{subfigure}[b]{0.24\textwidth}
                    \centering
                    \includegraphics[width=\linewidth]{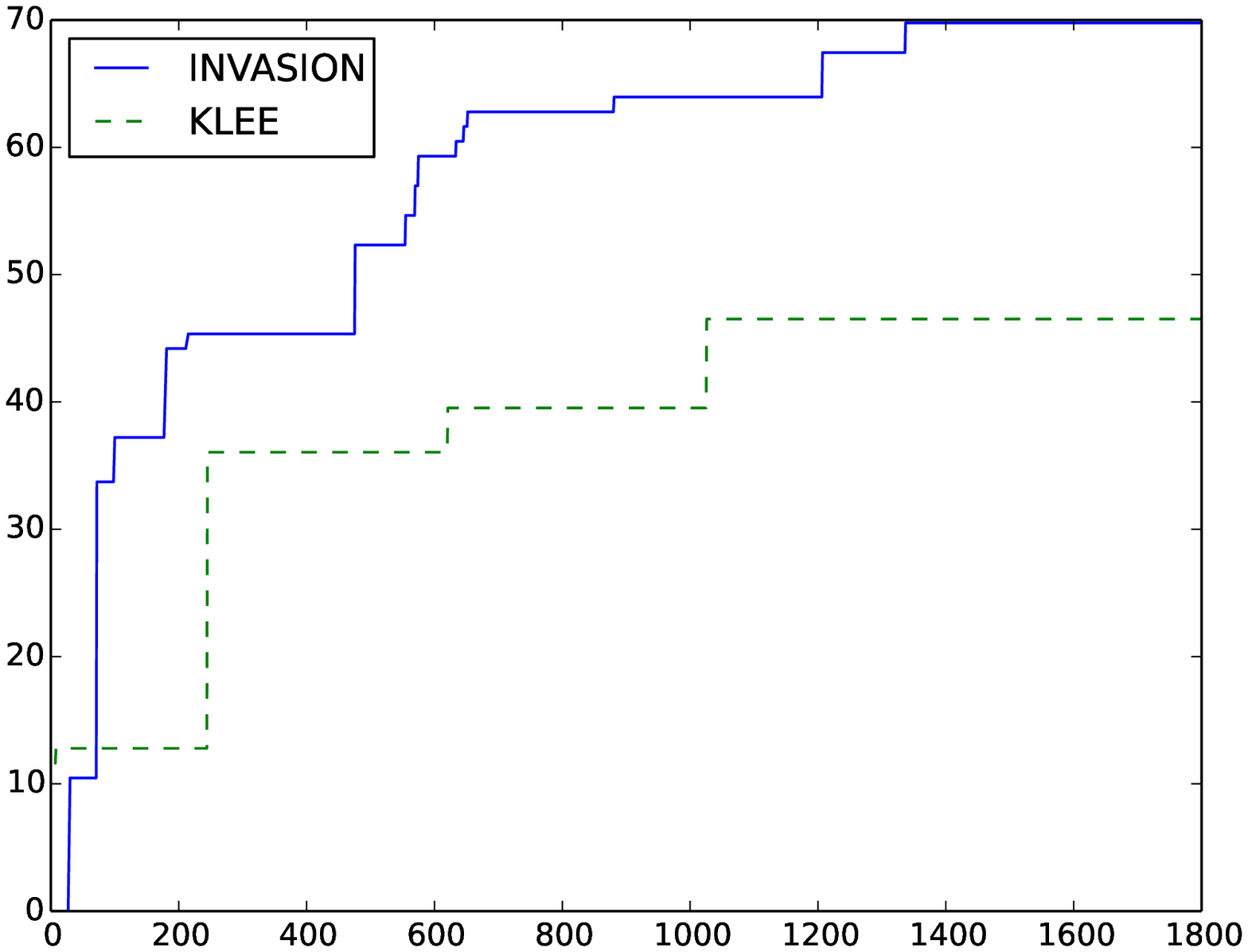}
                    \caption{Schedule2}
                    \label{fig:mouse}
            \end{subfigure}
            \begin{subfigure}[b]{0.24\textwidth}
                    \centering
                    \includegraphics[width=\linewidth]{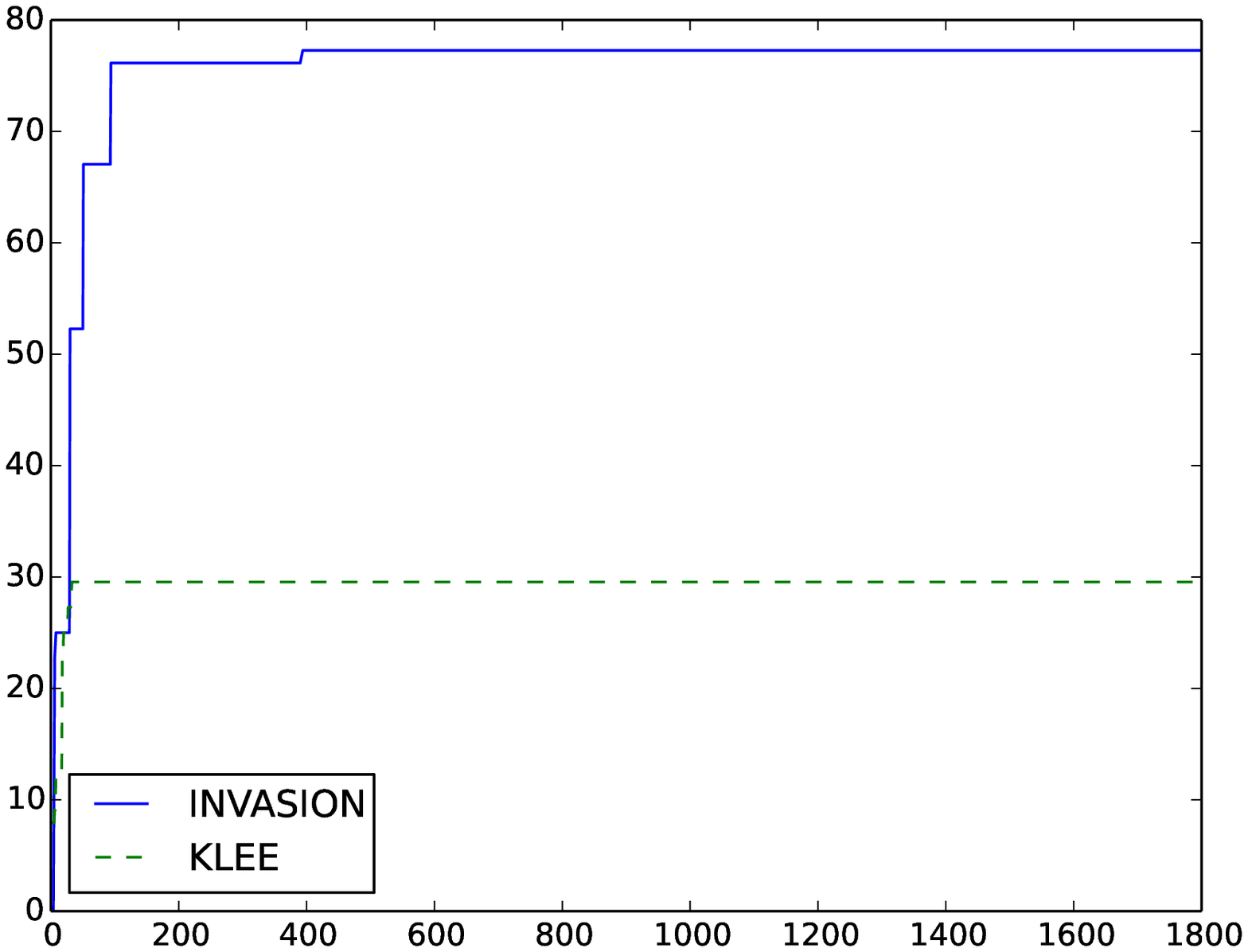}
                    \caption{Tot\_info}
                    \label{fig:mouse}
            \end{subfigure}
            \begin{subfigure}[b]{0.24\textwidth}
                    \centering
                    \includegraphics[width=\linewidth]{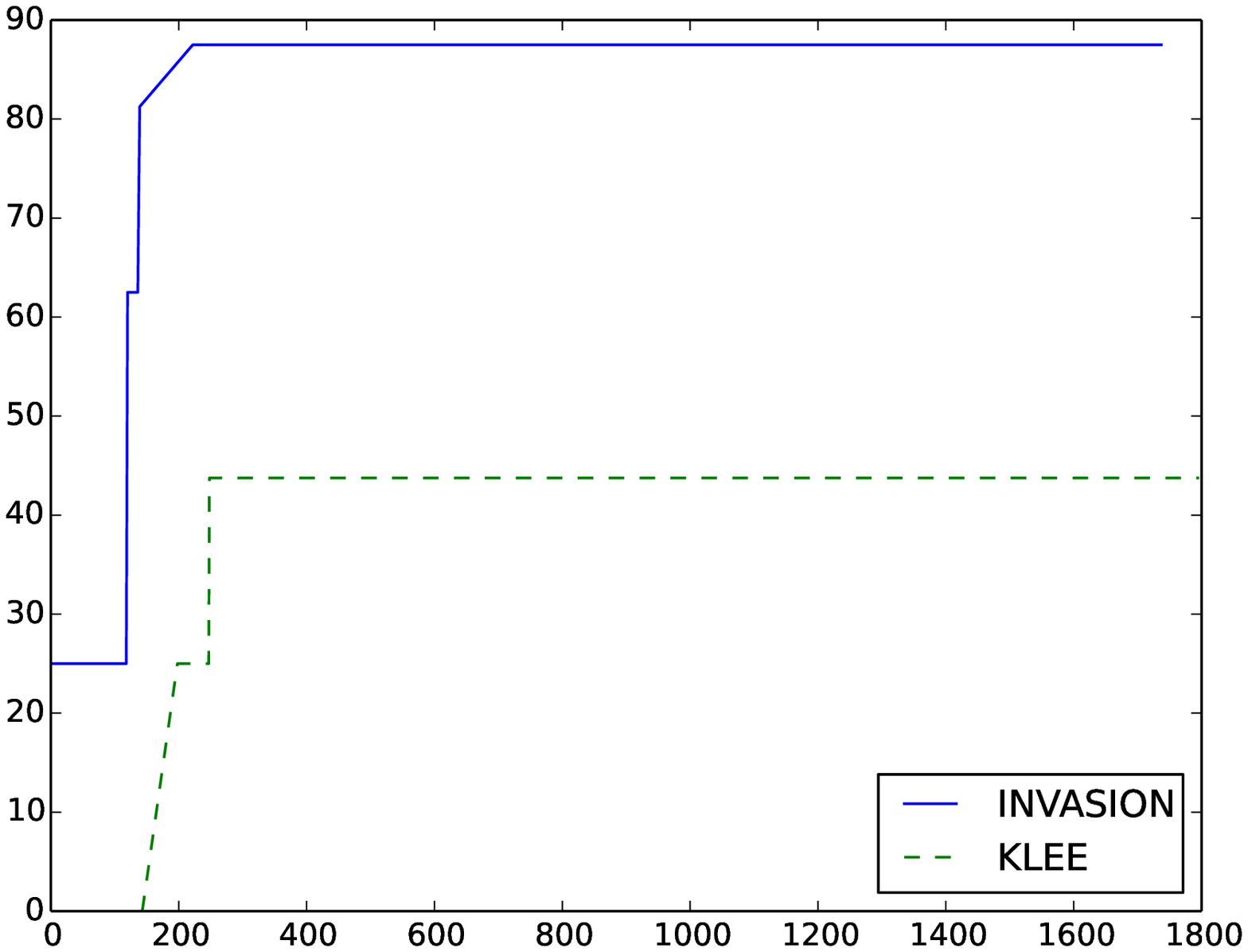}
                    \caption{FFT}
                    \label{fig:mouse}
            \end{subfigure}
            \begin{subfigure}[b]{0.24\textwidth}
                    \centering
                    \includegraphics[width=\linewidth]{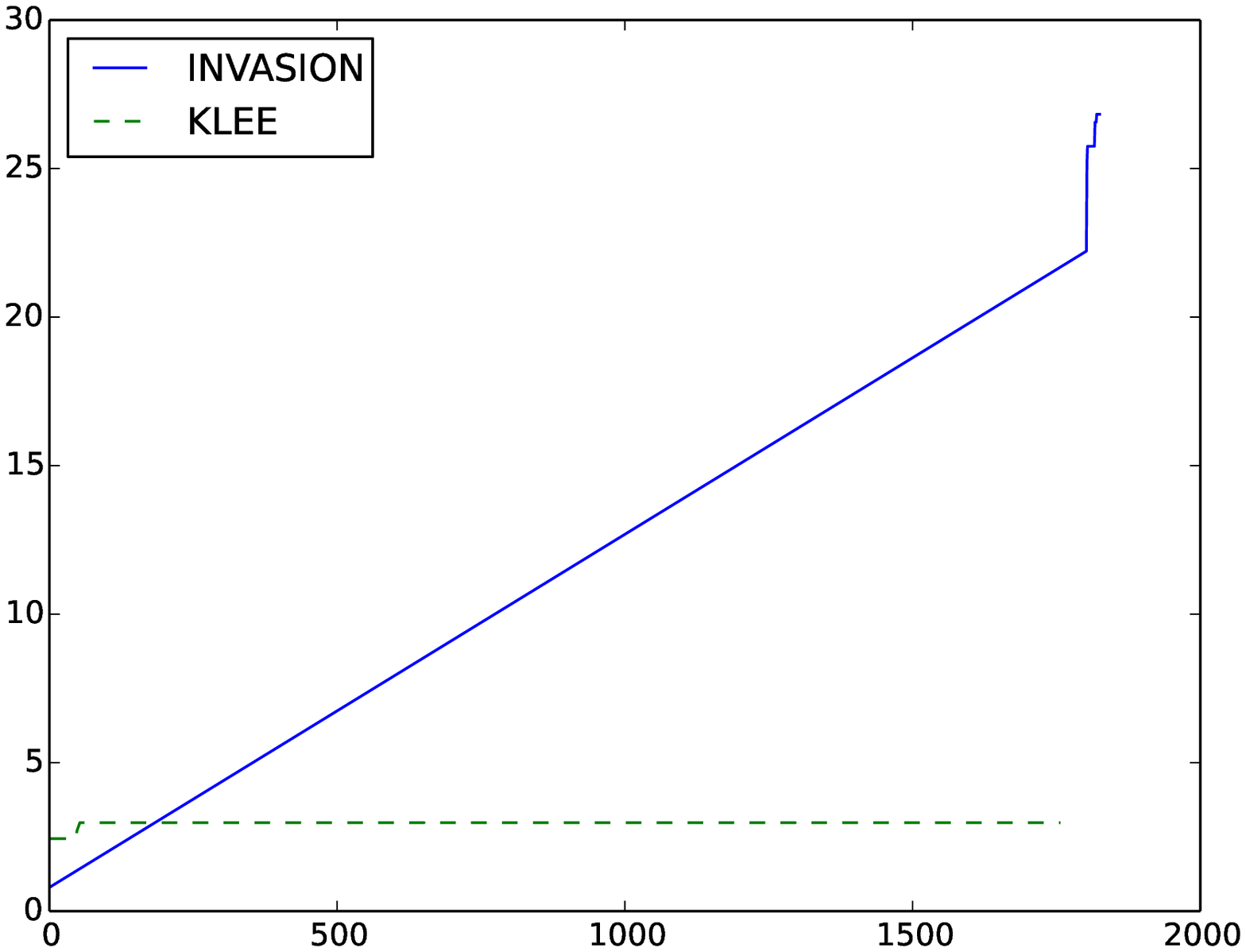}
                    \caption{TiffCmp}
                    \label{fig:mouse}
            \end{subfigure}
            \caption{Increase in coverage with respect to the time \textcolor{blue}{blue} is for \toolname and dotted \textcolor{green}{green} line is for KLEE}\label{fig:timeVsCov}
    \end{figure}

}

\section{Related Work}
The research community has witnessed an explosion of symbolic execution engines like KLEE~\cite{klee:2008}, Crest~\cite{crest}, EXE~\cite{exe}, Mayhem~\cite{mayhem} and S2E~\cite{s2e}. EXE does forking based symbolic execution and generates test cases which can reveal bugs in the program. Mayhem is a concolic execution engine, but it works on a binary program even if the binary does not contain debugging information. It does so by reasoning over the symbolic memory model. S2E is also a symbolic execution engine which works on binary programs. It does selective symbolic execution for reaching the targeted area as quickly as possible. 

Symbolic execution tools are widely used in testing and repairing a program. Symbolic execution has been used for program repair tasks in tools like Angelix~\cite{angelix:2016}, Semfix~\cite{semfix:2013}, Prophet~\cite{prophet}. Other Applications of symbolic execution engines are to find exploitable bugs in the systems. 


Amongst the proposals that attempt to handle formatted (structured) inputs, Caballero~\cite{cab} et.al uses protocol-level constraint-guided exploration for capturing inputs to expose vulnerabilities. 
They collect byte-level constraints, which are converted to protocol level by adding the protocol constraints. This reduces the number of paths explored compared to stream-based signatures. However, constraint sizes can explode for large protocol descriptions.  Wondracek~\cite{wan} et al. attempt reverse engineering protocol stacks via static analysis of binary programs to find loop exit points, after which, a taint analysis figures out responsible input variables. Then, they attempt to run the loop in a controlled manner. Our technique does not require any such analysis.

Majumdar~\cite{Majumdar} et. al designed a tool CESE, which can generate the input which is generated from a context-free grammar. They model the grammar of the input symbolically, and then feed these input to concolic execution engine. This improves the performance of the concolic execution engine. Their approach is similar to our work in that they translate the context-free grammar to logical constraints; however, we use a more expressive language and use the symbolic execution engine to translate our constraints to logic (by simply translating our program to a non-deterministic program, which is executed symbolically). Godefried~\cite{god} propose a similar technique for fuzzing over tokens from a context-free grammar. Their proposal is useful for testing tools like interpreters and parsers. They use a special solver that conjuncts the membership check in the input grammar with the path condition, to generate new feasible paths to be explored. 

Our work differs from these work in two primary ways: firstly, we propose our algorithm for a \textit{forking-based} symbolic execution engine, while their work is limited to an \textit{instrumentation-based} symbolic execution engine (we discuss the challenges and differences in \cref{intro}). Secondly, context-free languages are not a good specification language when a semantic interpretation of parts of the input is required (like in many programs, including Tot\_info, Tiff and Schedule benchmarks). We provide a more expressive language for such tasks. Finally, our approach is not targetted at any symbolic execution engine; it can be run with ease on any symbolic execution engine, for source and for binaries.

Medicherla~\cite{raghavan} et. al solved this problem of handling formatted inputs for improving the precision of static analysis (essentially, dataflow analysis~\cite{Aho}) file-processing programs. They perform data flow analysis on the program for finding nodes which are not directly affected by the input. However, our work is different than their as we target improving coverage of symbolic execution.


\section{Conclusion}
In this work, we motivate the use of input description languages to specify inputs for programs targeting symbolic execution. We propose a simple model, based on a finite-state automaton, to describe inputs and show it to be effective for programs requiring complex input specifications. We also build a tool, \toolname that instantiates our ideas. In the future, we are interested in the problem of automatically learning our input description and scheduling the fragments of the input specification in a manner such that inputs which are likely to induce faults are prioritized.

\bibliographystyle{splncs}
\bibliography{ref}
\end{document}